\documentclass[rnote]{myaa}
\usepackage{psfig}

\def\la{\;
\raise0.3ex\hbox{$<$\kern-0.75em\raise-1.1ex\hbox{$\sim$}}\; }
\def\ga{\;
\raise0.3ex\hbox{$>$\kern-0.75em\raise-1.1ex\hbox{$\sim$}}\; }

\newcommand{\zabs}{$z_{\rm abs}\,$}
\newcommand{\zem}{$z_{\rm em}\,$}

\newcommand{\kms}{km~s$^{-1}\,$}
\newcommand{\ms}{m~s$^{-1}\,$}
\newcommand{\cm}{cm$^{-2}\,$}

\newcommand{\daa}{$\Delta\alpha/\alpha\,$}

\idline{0}{1}
\begin{document}

\title{A new measure of $\Delta\alpha/\alpha$ at redshift $z = 1.84$
from very high resolution spectra of \object{Q 1101--264}\thanks{Based on 
observations performed at the VLT Kueyen telescope (ESO, Paranal, Chile), 
the ESO programme No.~076.A-0463
}
}
\author{
S. A. Levshakov\inst{1}\thanks{On leave from the Ioffe
Physico-Technical Institute, St. Petersburg, Russia}
\and
P. Molaro\inst{2,3}
\and
S. Lopez\inst{4}
\and
S. D'Odorico\inst{5}
\and
M. Centuri\'on\inst{2}
\and
P. Bonifacio\inst{2,3}
\and\\
I. I. Agafonova\inst{1}\thanks{On leave from the Ioffe
Physico-Technical Institute, St. Petersburg, Russia}
\and
D. Reimers\inst{1}
}
\offprints{S.~A.~Levshakov
\protect \\lev@astro.ioffe.rssi.ru}
\institute{
Hamburger Sternwarte, Universit\"at Hamburg,
Gojenbergsweg 112, D-21029 Hamburg, Germany
\and
Osservatorio Astronomico di Trieste, Via G. B. Tiepolo 11,
34131 Trieste, Italy
\and
Observatoire de Paris 61, avenue de l'Observatoire, 75014 Paris, France
\and 
Departamento de Astronom\'ia, Universidad de Chile,
Casilla 36-D, Santiago, Chile
\and
European Southern Observatory, Karl-Schwarzschild-Strasse 2,
D-85748 Garching bei M\"unchen, Germany
}
\date{Received 00  / Accepted 00 }
\abstract{}
{We probe the evolution of the fine-structure constant $\alpha$ with
cosmic time.
}
{Accurate positions of the \ion{Fe}{ii} lines $\lambda1608$,
$\lambda2382$, and $\lambda2600$ are measured in the \zabs = 1.84 
absorption system from a
high-resolution ($FWHM \sim 3.8$ \kms) and high signal-to-noise
(S/N $\ga 100$) spectrum of the quasar
\object{Q 1101--264} (\zem = 2.15, $V = 16.0$),
integrated for 15.4 hours.
The Single Ion Differential $\alpha$ Measurement (SIDAM) procedure
and the $\Delta \chi^2$ method are used to set constraints on \daa.
}
{
We have found a relative radial velocity shift between the $\lambda1608$
and $\lambda\lambda2382,2600$ lines of 
$\Delta v = -180\pm85$ m~s$^{-1}$ (both random 
and systematic errors are included),
which, if real, would correspond to
\daa = $(5.4\pm2.5)\times10^{-6}$ ($1\sigma$ C.L.).
Considering the strong implications of a such variability, additional
observations with comparable accuracy at redshift $z \sim 1.8$  are
required to confirm this result.
}
{}

\authorrunning{S. A. Levshakov et al.}
\titlerunning{$\Delta\alpha/\alpha$ at $z = 1.84$ }
\maketitle

\section{Introduction}

The current cosmological model ($\Lambda$CDM, for $\Lambda$ Cold Dark Matter) 
assumes that our universe is dominated by dark matter 
and dark energy. 
This model is based on a series of independent
observations such as 
the luminosity distances of high-redshift type Ia supernovae (SNe Ia),
the cosmic microwave background temperature anisotropy, 
the large scale distribution
of galaxies, and the integrated Sachs-Wolfe effect
(for a review, see Copeland et al. 2006).

The observational data on SNe Ia suggest that
the kinematics of the expansion of the universe, 
decelerating at redshift $z > 1$, has
changed to the accelerating phase at lower $z$. 
This late-time acceleration of the universe is regarded as
evidence for a non-zero cosmological constant $\Lambda$ or
for the existence of dark 
energy with negative pressure.

The nature of the dark energy is still unknown. At the moment,
a variety of models is being considered, from scalar fields with
different potentials to extra dimensions ($D > 4$) of physical
space-time (e.g., Dvali et al. 2000).
In many models, the cosmological evolution of dark energy is accompanied by
variations in coupling constants, the fine-structure constant $\alpha$
and the electron-to-proton mass ratio.
This means that 
the evolution of dark energy~--- if it indeed occurs~---
can be traced 
through the measurements of the coupling constants changes
in course of cosmic time (e.g., Avelino et al. 2006).
Theories predict different behavior of dark energy, from slow-rolling to
oscillating 
(e.g., Liddle \&  Lyth 2000; Fujii \& Maeda 2003). 
It is clear that in order to study the dynamics of dark energy the
fundamental constants must be measured at every redshift $z$ with highest
accuracy. 

The value of \daa = $(\alpha_z - \alpha)/\alpha$ at a redshift $z$
can be estimated from the radial velocity shifts of the line
positions of different ions. Theoretical basics of the method called
`Many Multiplet (MM) method' are given in Webb et al. (1999) and
Dzuba et al. (2002).
Recently obtained results are as follows.

Murphy et al. (2004) claim that  \daa = $-5.7\pm1.1$ ppm 
(ppm stands for parts per million, $10^{-6}$)
in the redshift range $0.2 < z < 4.2$, i.e.
$\alpha$ was lower in the past. 
This result was produced by averaging over
143 absorption systems detected in HIRES/Keck telescope spectra of QSOs.
However, Chand et al. (2004) report \daa = $-0.6\pm0.6$ ppm
for $0.4 < z < 2.3$ from the analysis of
23 absorption systems identified in the UVES/VLT spectra of QSOs.
It should be emphasized that 
\begin{enumerate}
\item[--]
both the Murphy et al. and Chand et al.
estimates are obtained by averaging over many absorption systems
with different redshifts, i.e. the accuracy of
\daa\ at individual redshifts is considerably lower than the
reported values; 
\item[--]
\ion{Mg}{ii} and \ion{Fe}{ii} ions~--- the main elements of the most accurate low-$z$
sample by Murphy et al.~--- cannot be formed in the same volume of the intervening
metal absorbers as shown from spectral observations of GRBs (Hao et al. 2006; Porciani et al.
2007).
\end{enumerate}

Recent laboratory restriction on the rate of fractional
variation of the fine-structure constant
of $(-2.7\pm2.6)\times10^{-15}$ yr$^{-1}$
(Cing\"oz et al. 2007), which is based on highly sensitive to temporal
variation of $\alpha$ transition frequencies in two isotopes of dysprosium
(Dy), indicates no significant variation at present time $(z = 0)$.
Revision of the fission products of a natural reactor in Oklo ($z = 0.4$)
also indicates no variation at the level of
$1.2\times10^{-17}$ yr$^{-1}$\
(Gould et al. 2006).

In order to improve the accuracy of individual \daa values
measured from quasar absorption-line spectra,
we developed a modification of the MM method called
`Single Ion Differential $\alpha$ Measurement, SIDAM' 
(Levshakov 2004; Quast et al. 2004; 
Levshakov et al. 2005, hereafter Paper~I; 
Levshakov et al. 2006, hereafter Paper~II).
Based on the analysis of only
one element (singly charged iron, \ion{Fe}{ii}), it is free
from the influence of small Doppler shifts between different
elements caused by ionization inhomogeneities within the absorber, 
and by unknown isotopic abundances.
It is also free from different line profile distortions due to
finite correlation length effects in turbulent velocity
fields which are sensitive to the ratio of the rms turbulent velocity
to the thermal width of the absorbing ions (Levshakov \& Kegel 1997;
Levshakov et al. 1997).

Using SIDAM, we have reached the accuracy of \daa 
for one absorption system at the level of $\sim 10^{-6}$, i.e.
as high as the accuracy of the mentioned above
values averaged over many absorption systems.
Our previous results give the following constraints on \daa.

At  \zabs = 1.15 towards \object{HE 0515--4414},
\daa = $-0.4\pm1.9$ ppm at \zabs = 1.15 (Quast et al. 2004). 
Later on, we improved the accuracy of this estimation and obtained
\daa = $-0.07\pm0.84$ ppm (Paper~II).
This utmost accuracy was achieved due to unique favorable conditions
such as the brightness of the quasar and the strength of all \ion{Fe}{ii}
lines including $\lambda1608$ \AA.
In both cases the same archive data were used:
$FWHM \simeq 5.6$ \kms in the blue 
and $\simeq 4.8-5.5$ \kms in the red. 
A consistent restriction on \daa in this system was obtained
independently by Chand et al. (2006), 
\daa = $0.5\pm2.4$ ppm, who used 
the high resolution pressure and temperature stabilized spectrograph
HARPS mounted on the ESO 3.6~m telescope at the La Silla observatory.

Further, using
the VLT/UVES archive spectra of \object{Q 1101--264} with
the resolution $FWHM \simeq 6.0$ \kms in the blue ($\lambda \sim 4500$
\AA), and $\simeq 5.4$ \kms in the red ($\lambda \sim 7000$ \AA) 
we obtained \daa = $2.4\pm3.8$ ppm at
\zabs = 1.839 (Paper~I).

Recently this quasar was re-observed with higher spectral resolution.
In the present paper we describe the new measurement of \daa
at \zabs = 1.839 
which allowed us to enhance the accuracy of \daa\ by a factor 1.5.

\section{Observations and Data Reduction}

The observations of \object{Q 1101--264} were acquired with the 
VLT UV-Visual Echelle
Spectrograph (UVES) on the nights of 2006 February 21, 22, and 23.
The selected exposures are listed in Table~1.
The spectra were recorded with a dichroic filter which allows
to work with the blue and red UVES arms simultaneously as with 
two independent spectrographs, 
and the CCDs were read out  unbinned.
The standard settings at central wavelengths 
$\lambda$437 nm and $\lambda$860 nm
were used for the blue and red arms respectively (marked by symbols `b' 
and `r' in Col.~1 of Table~1).  
The slit was oriented along the parallactic angle
and the width set at 0.5 arcsec, with DIMM values 
ranging from 0.5 to 1.1 arcsec during the run. 

Calibration exposures were taken immediately after 
scientific exposures to minimize the influence of
changing ambient weather conditions which may cause
different velocity offsets in the lamp and QSO spectra if they were
not obtained closely in time (Paper~II).
Variations in temperature can induce color effects in the spectra
since they act differently on the different cross dispersers. 

\begin{table}[t!]
\centering
\caption{UVES/VLT observations of \object{Q 1101--264} 
(ESO programme 076.A-0463) }
\label{tbl-1}
\begin{tabular}{cccrc}
\hline
\noalign{\smallskip}
\multicolumn{1}{l}{\footnotesize Exp.} & 
{\footnotesize Date} & 
{\footnotesize Time} & \multicolumn{1}{c}{\footnotesize Exp.} & 
{\footnotesize S/N per}\\[-2pt] 
{\footnotesize No.}  & {\footnotesize y-m-d} & {\footnotesize h:m:s} & 
\multicolumn{1}{c}{\footnotesize s} & {\footnotesize pixel} \\[-2pt] 
{\scriptsize (1)} & {\scriptsize (2)} & 
{\scriptsize (3)} & \multicolumn{1}{c}{\scriptsize (4)} & {\scriptsize (5)} \\
\hline
\noalign{\smallskip}
\noalign{\smallskip}
{\scriptsize 1b}&{\scriptsize 2006-02-21}&
{\scriptsize 07:39:26}&{\scriptsize 6400}&{\scriptsize 30}\\[-2pt]
{\scriptsize 1r}&{\scriptsize 2006-02-21}&
{\scriptsize 07:39:22}&{\scriptsize 5400}&{\scriptsize  32,34}\\
\noalign{\smallskip}
{\scriptsize 2b}&{\scriptsize 2006-02-22}&{\scriptsize 02:28:57}&
{\scriptsize 5400}&{\scriptsize 31}\\[-2pt]
{\scriptsize 2r}&{\scriptsize 2006-02-22}&{\scriptsize 02:28:52}&
{\scriptsize 5400}&{\scriptsize  39,37}\\
\noalign{\smallskip}
{\scriptsize 3b}&{\scriptsize 2006-02-22}&{\scriptsize 04:04:47}&
{\scriptsize 4953}&{\scriptsize 30}\\[-2pt]
{\scriptsize 3r}&{\scriptsize 2006-02-22}&{\scriptsize 04:04:43}&
{\scriptsize 4959}&{\scriptsize  44,44}\\
\noalign{\smallskip}
{\scriptsize 4b}&{\scriptsize 2006-02-22}&{\scriptsize 05:55:21}&
{\scriptsize 5400}&{\scriptsize 34} \\[-2pt]
{\scriptsize 4r}&{\scriptsize 2006-02-22}&{\scriptsize 05:55:17}&
{\scriptsize 5400}&{\scriptsize 47,49}\\
\noalign{\smallskip}
{\scriptsize 5b}&{\scriptsize 2006-02-22}&{\scriptsize 07:30:28}&
{\scriptsize 6708}&{\scriptsize 38} \\[-2pt]
{\scriptsize 5r}&{\scriptsize 2006-02-22}&{\scriptsize 07:30:25}&
{\scriptsize 6712}&{\scriptsize 52,48}\\
\noalign{\smallskip}
{\scriptsize 6b}&{\scriptsize 2006-02-23}&{\scriptsize 01:29:26}&
{\scriptsize 6300}&{\scriptsize 26} \\[-2pt]
{\scriptsize 6r}&{\scriptsize 2006-02-23}&{\scriptsize 01:29:25}&
{\scriptsize 6300}&{\scriptsize 36,38}\\
\noalign{\smallskip}
{\scriptsize 7b}&{\scriptsize 2006-02-23}&{\scriptsize 03:19:23}&
{\scriptsize 7200}&{\scriptsize 35} \\[-2pt]
{\scriptsize 7r}&{\scriptsize 2006-02-23}&{\scriptsize 03:19:22}&
{\scriptsize 7200}&{\scriptsize 46,48}\\
\noalign{\smallskip}
{\scriptsize 8b}&{\scriptsize 2006-02-23}&{\scriptsize 05:45:15}&
{\scriptsize 5400}&{\scriptsize 37} \\[-2pt]
{\scriptsize 8r}&{\scriptsize 2006-02-23}&{\scriptsize 05:45:10}&
{\scriptsize 5400}&{\scriptsize 41,46}\\
\noalign{\smallskip}
{\scriptsize 9b}&{\scriptsize 2006-02-23}&{\scriptsize 07:19:40}&
{\scriptsize 7500}&{\scriptsize 40} \\[-2pt]
{\scriptsize 9r}&{\scriptsize 2006-02-23}&{\scriptsize 07:19:34}&
{\scriptsize 7500}&{\scriptsize 52,53}\\
\noalign{\smallskip}
\hline
\noalign{\smallskip}
\multicolumn{5}{l}{\footnotesize Note: in Col.~1, `b' and `r' denote
the blue-arm (echelle}\\[-2pt] 
\multicolumn{5}{l}{\footnotesize order 102) 
and red-arm (echelle orders 90 and 83) }\\[-2pt]
\multicolumn{5}{l}{\footnotesize  exposures;
calibration lamp exposures were taken }\\[-2pt]
\multicolumn{5}{l}{\footnotesize 
immediately after each science exposure; 
the pixel}\\[-2pt]
\multicolumn{5}{l}{\footnotesize sizes close to the
\ion{Fe}{ii} lines, where the S/N has been}\\[-2pt] 
\multicolumn{5}{l}{\footnotesize computed, are 25 m\AA\
($\lambda1608$), 30 m\AA\ ($\lambda2382$), and}\\[-2pt]
\multicolumn{5}{l}{\footnotesize 27 m\AA\  ($\lambda2600$).}
\end{tabular}
\end{table}

To control possible systematics in radial velocities, the thermal and
pressure stabilities were monitored.
Variations of the ambient temperature during science exposures (the
difference $\Delta T$ between the beginning and the end) were always
less than 0.1~K. This is also the maximum variation occurring between the
science exposure and the calibration lamp. The pressure variation
$\Delta P$ reached 0.6 mbar in one case (No.~4, Table~1), in all other
cases the variations of $\Delta P$ did not exceed 0.3 mbar.
The estimations of Kaufer et al. (2004) are of 50 \ms\ for 
$\Delta T = 0.3$~K and $\Delta P = 1$ mbar, so the observed 
differences should not affect significantly the radial velocities. 

From the blue frame we used only order 102, and from 
the red~--- orders 90 and 83 (Col.~5),
where \ion{Fe}{ii} lines with, respectively, 
$\lambda = 1608$, 2382 and 2600 \AA\ are observed. 
The selected lines are located close 
to the central regions of the corresponding
echelle orders.  
This minimizes possible distortions of the line profiles
caused by the decreasing spectral
sensitivity at the edges of echelle orders.
The signal-to-noise ratios (S/N) per pixel (25 m\AA,
30 m\AA\ and 27 m\AA\ at the positions of $\lambda1608$,
$\lambda2382$ and $\lambda2600$, respectively) 
for the individual exposures
and  individual orders are listed in Col.~6.
Cols.~2, 3 and 4 give, respectively, the date of observation,
the beginning of the exposure, and the total exposure time.

The spectra were reduced using the {\tt UVES} context within 
the ESO MIDAS data reduction software. 
Master flat-field and master bias
frames were obtained by computing the median of five flat-field
and bias frames taken in the afternoon. 
The master bias was subtracted
from each science frame; subsequently the background was subtracted
by fitting a two-dimensional spline to points in the inter-order
region, where the background was estimated as the median number
of counts over a box 51 pixels wide in the dispersion direction   
and 5 pixels wide in the cross-dispersion direction.
The background subtracted science frame was divided pixel-by-pixel
by the normalized master flat-field.
The spectrum was extracted using an average extraction and
an optimal extraction. The S/N ratio proved to be higher
in the average extraction, therefore this was used
for the analysis.
The wavelength calibration was derived from the spectrum of a
ThAr hollow cathode lamp. We used the Los Alamos table of ThAr lines
(Palmer \& Engleman 1983; de Cuyper \& Hensberge 1988) and the 1D option,
by which a 4th order polynomial is fit to the measured
positions of arc lines in each echelle order.
The wavelength calibration was applied to the extracted science
spectrum both on a pixel by pixel basis, and rebinning the
extracted spectrum to a constant wavelength step.
We did not observe any systematic shift in the line centroids
introduced by the rebinning, therefore we used the rebinned spectra,
which can be readily co-added.
We kept a very small tolerance in the fitting in order to reject
several faint lines with lower accuracy (about 150 \ms)
still preserving somewhat 20 lines for each order, normally
strong lines with about 15 \ms\ accuracy.
Finally, as a cross-check, we performed an independent reduction
of the spectra using a 2D wavelength calibration, by which a
2D polynomial dispersion relation of the form $f(m,\lambda)$,
where $m$ is the absolute order number and $\lambda$ is the wavelength,
is fitted to the entire echellogram of the ThAr lamp (Lopez et al. 2005). 
The residuals of the calibrations in both approaches
were rather small corresponding to the rms errors of the wavelength scale
calibration of about 20 \ms.
The observed wavelength scale of each spectrum was transformed into vacuum,
heliocentric wavelength scale (Edl\'en 1966). 
The resulting spectral resolution as measured from the ThAr emission
lines was $FWHM \simeq 3.95$ \kms in the blue ($\lambda \sim 4500$ \AA)
and $\simeq 3.75$ \kms in the red ($\lambda \sim 7000$ \AA).
Thus, the resolution of these new spectra is significantly higher
than in the previous observations (6 and 5.4 \kms respectively, see
Paper I).

\begin{figure}[t]
\vspace{0.0cm}
\hspace{-0.5cm}\psfig{figure=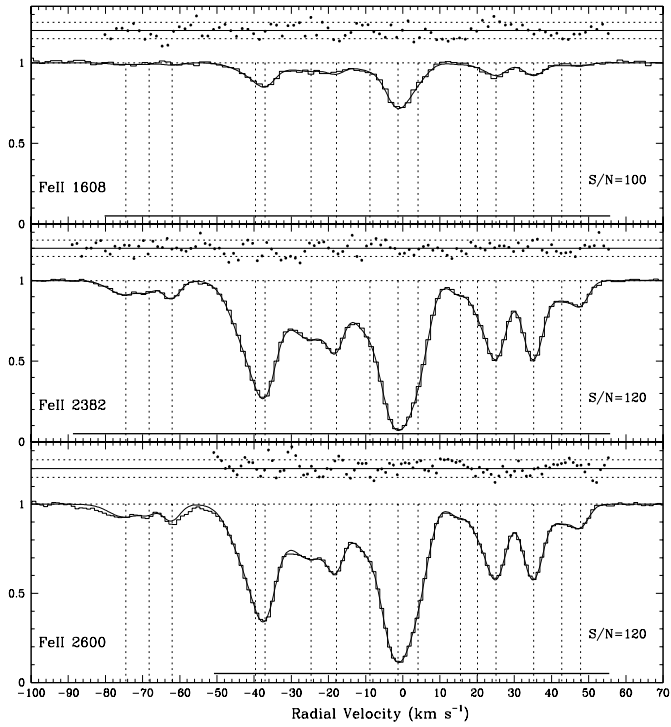,height=10.0cm,width=9.5cm}
\vspace{-0.6cm}
\caption[]{Combined absorption-line spectra of \ion{Fe}{ii}
associated with the $z = 1.84$ damped Ly$\alpha$ system towards
\object{Q 1101--264} (normalized intensities are shown by histograms).
The zero radial velocity is fixed at \zabs = 1.838911.
The synthetic profiles are over-plotted by the smooth curves.  
The normalized residuals,
$({\cal F}^{cal}_i-{\cal F}^{obs}_i)/\sigma_i$,
are shown by dots (horizontal dotted lines restrict the $1\sigma$ errors). 
The dotted vertical lines mark positions of the 
sub-components. 
Bold horizontal lines mark pixels included in the
optimization procedure. 
The ranges at $v < -50$ \kms and 
at $v \simeq -30$ \kms in the \ion{Fe}{ii} $\lambda2600$ profile
are blended with weak telluric lines. 
The normalized $\chi^2_\nu = 0.901\, (\nu = 257)$. 
}
\label{fig1}
\end{figure}

\begin{figure}[t]
\vspace{0.0cm}
\hspace{-0.5cm}\psfig{figure=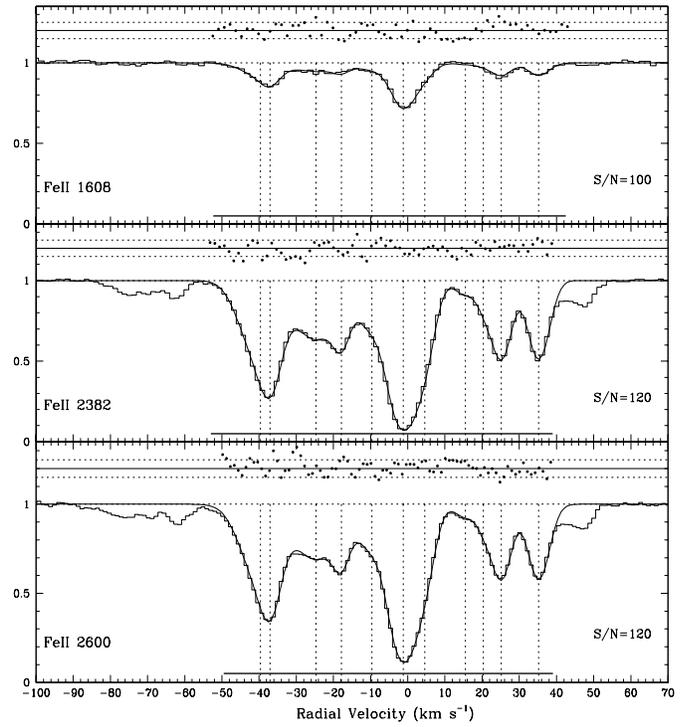,height=10.0cm,width=9.5cm}
\vspace{-0.6cm}
\caption[]{Same as Fig.~1 but for the 11-component model
fitted to the central part of the \ion{Fe}{ii} lines.
The normalized $\chi^2_\nu = 0.720\, (\nu = 188)$.
}
\label{fig2}
\end{figure}

\begin{figure}[t]
\vspace{0.0cm}
\hspace{-0.5cm}\psfig{figure=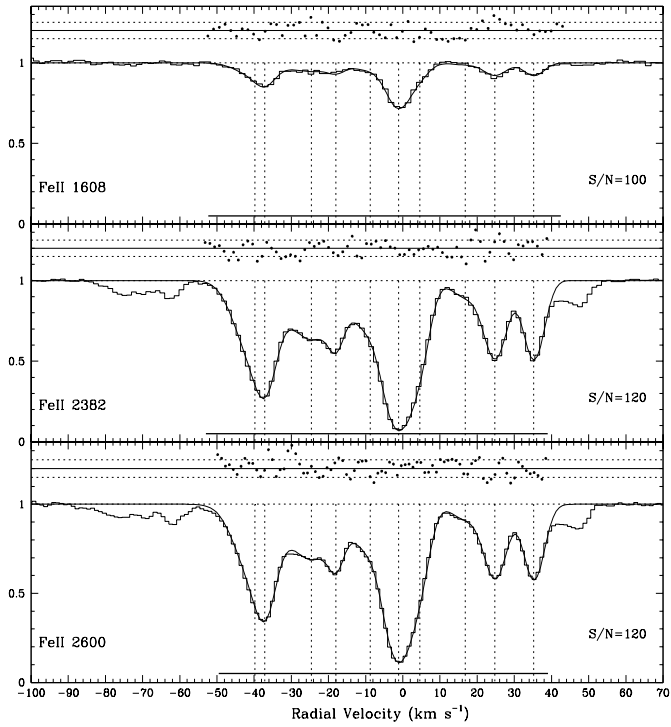,height=10.0cm,width=9.5cm}
\vspace{-0.6cm}
\caption[]{Same as Fig.~2 but for the 10-component model.
The normalized $\chi^2_\nu = 0.871\, (\nu = 191)$.
}
\label{fig3}
\end{figure}

\begin{figure}[t]
\vspace{0.0cm}
\hspace{-0.3cm}\psfig{figure=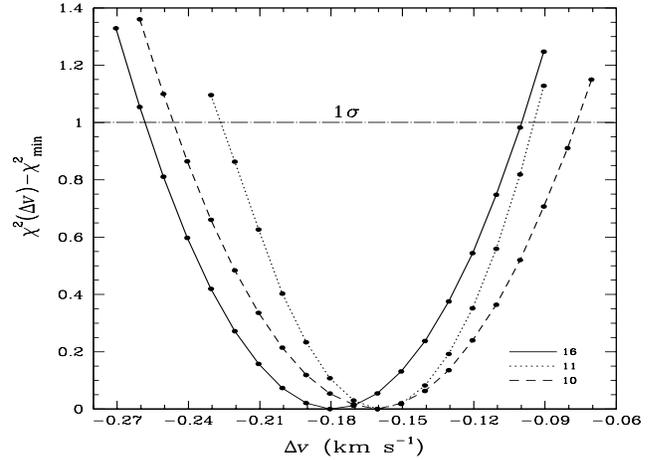,height=7.0cm,width=9.5cm}
\vspace{-0.9cm}
\caption[]{$\chi^2$ as a function of the velocity difference $\Delta v$
between the \ion{Fe}{ii} $\lambda1608$ and $\lambda\lambda2382, 2600$
lines for the 16-, 11-, and 10-component models which are shown, respectively,
in Figs.~1, 2, and 3. The corresponding $\chi^2_{\rm min}$ values
are equal to 231.631, 135.314, and 166.397.
The minima of the curves
give the most probable values of \daa\ (see eq.[1]):
5.4 ppm (the 16-component model)
and 4.8 ppm (the 11- and 10-componet models).
The $1\sigma$ confidence level is determined by $\Delta \chi^2 = 1$
which gives $\sigma_{\Delta v} = 0.080$ \kms,
0.065 \kms, and 0.085 \kms
or
$\sigma_{\Delta \alpha/\alpha} = 2.4$ ppm, 1.9 ppm, and  2.5 ppm for
the 16-, 11-, and 10-component models, respectively.
}
\label{fig4}
\end{figure}

\section{Estimate of \daa}

The spectroscopic measurability of \daa is based on the fact that the
energy of each line transition depends individually on a change in
$\alpha$ (e.g., Webb et al. 1999).
It means that the relative change of the frequency $\omega_0$
due to varying $\alpha$ is proportional to the so-called 
sensitivity coefficient ${\cal Q} = q/\omega_0$ (Paper~I). 
The $q$-values
for the resonance UV transitions in \ion{Fe}{ii} are calculated by
Dzuba et al. (2002).  The corresponding sensitivity coefficients 
are listed in Table~5 in Paper~II.  
The rest frame wavelengths for some \ion{Fe}{ii} transitions
have been re-measured recently 
with higher accuracy by Aldenius et al. (2006).
For our present work we need two of them: $2382.7641\pm0.0001$ \AA,
and $2600.1722\pm0.0001$ \AA. The line from the 8th UV multiplet
$1608.45080\pm0.00008$ \AA\ 
in Table~5 is registered to the calibration scale defined in
Norl\'en (1973), whereas the lines $\lambda2382$ and $\lambda2600$
from Aldenius et al. are calibrated in the scale of Whaling et al. (1995).
The two scales differ by a factor of about $7\times10^{-8}$ in 
frequency space. Thus, a corrected wavelength of the $\lambda1608$ line
is 1608.45069 \AA. 
Other atomic data are taken 
as they are in Table~5. 

The value of \daa itself depends on a proper interpretation of
measured relative radial velocity shifts, $\Delta v$, between
lines with different sensitivity coefficients,
$\Delta {\cal Q}$. 
It can be shown that
in linear approximation ($|\Delta\alpha/\alpha|\ll1$) 
these quantities are related as (Paper~II):
\begin{equation}
\frac{\Delta\alpha}{\alpha} = \frac{(v_2 - v_1)}
{2\,c\,({\cal Q}_1 - {\cal Q}_2)} = 
\frac{\Delta v}{2\,c\,\Delta {\cal Q}}\ .
\label{E1}
\end{equation}
Here index `1' is assigned to the line $\lambda1608$, while
index `2' marks one of the other \ion{Fe}{ii} lines ($\lambda2382$ 
or $\lambda2600$). 

It is important that $\Delta {\cal Q}$ must be large enough to provide
smaller uncertainty in \daa. With ${\cal Q}_1 = -0.021$ and
${\cal Q}_2 = 0.035$ for both $\lambda2382$ and $\lambda2600$ lines,
we have $\Delta {\cal Q} = -0.056$. Note that 
$|\Delta {\cal Q}| = 0.056$ is almost two times larger than 
 $|\Delta {\cal Q}| \simeq 0.03$ which is used in the standard MM method
when \ion{Mg}{ii} $\lambda\lambda2796, 2803$ and 
\ion{Fe}{ii} $\lambda\lambda2344, 2374, 2382, 2586, 2600$ transitions
are compared\footnote{For \ion{Mg}{ii},
${\cal Q}_{2796} = 0.0059$, and ${\cal Q}_{2803} = 0.0034$.
For \ion{Fe}{ii}, ${\cal Q}_{2344} = 0.028$, 
${\cal Q}_{2374} = 0.038$, and ${\cal Q}_{2586} = 0.039$.}.
It is also important that the \ion{Fe}{ii} transitions at
$\lambda1608$ and $\lambda\lambda2382, 2600$ used in the present study
have sensitivity coefficients of different signs which helps
to distinguish the influence of hidden blends on the line position
measurements (Levshakov \& D'Odorico 1995).

The individual vacuum-barycentric spectra were re-sampled to
equidistant wavelength grids using smallest pixel sizes in the blue
and red exposures and applying linear interpolation.
The resulting normalized and co-added spectra from 9 exposures (Table~1) are
shown in Fig.~1.
A weighted mean signal-to-noise ratio per pixel of S/N~= 100 and
120 was achieved in the final spectrum at $\lambda \sim 4566$ \AA\
and $\lambda \sim 6765, 7382$ \AA, respectively. 
Two independent data reduction procedures resulted in almost
identical co-added spectra.
In contrast to the archive data used in
Paper~I which we failed to fit self-consistently, this time we found
a good fit of \ion{Fe}{ii} profiles to a model, 
the normalized $\chi^2$ per degree of freedom equals
$\chi^2_\nu = 0.901$ ($\nu = 257$).
The computational procedure was the same as in Paper~II.
The optimal solution was found for 
the 16-component model shown in
Fig.~1 by the smooth curves (the positions of the sub-components are
marked by dotted lines). 
The model parameters are given in Table~2.
Since the ${\cal Q}$ values for $\lambda2382$
and $\lambda2600$ are equal, their relative velocity shift 
characterizes the goodness of wavelength calibrations of the echelle 
orders 90 and 83. The revealed velocity shift of 20 \ms is 
comparable with 
the uncertainty range estimated from the ThAr lines. So, in what follows
we consider $\lambda2382$
and $\lambda2600$ as having the same radial velocity,
and calculate $\Delta v$ between this velocity
and that of the line $\lambda1608$.

To demonstrate that our result is not sensitive to the
number of subcomponents, we have analyzed different models.
The 16-component model is chosen
as one which covers the whole \ion{Fe}{ii} profile in the range
from $-90$ \kms\ to 55 \kms\ (Fig.~1) and
provides $\chi^2_\nu \la 1$ with minimum number of
components. The other models are shown in Figs.~2 and 3 with the
corresponding parameters listed in Table~2.
The 11-component model in Fig.~2
ignores the weak absorption in the
wings of the \ion{Fe}{ii} lines at $-90$ \kms $\la v \la -55$ \kms\
and 40 \kms $\la v \la 55$ \kms.
In the model shown in Fig.~3 the number of components is reduced
to 10 which is found to be the lower limit since models with
9 components 
cannot be fitted adequately to the \ion{Fe}{ii} lines in the range
$-55$ \kms $\la v \la 40$ \kms\ with $\chi^2_\nu$ less than 3.
We note that all of the analyzed models
show similar radial velocity shifts between the $\lambda1608$
and $\lambda2383/2600$ lines.
The comparison between the theoretical and observed \ion{Fe}{ii}
profiles was quantified by the $\chi^2$ function 
(defined by eq.[2] in Paper~II).
To find the most probable value of $\Delta v$, we fit the absorption lines
with a fixed $\Delta v$, changing $\Delta v$ in the interval from
$-270$ \ms to $-90$ \ms 
in steps of 10 \ms (see Fig.~4). For each
$\Delta v$, the strengths of the sub-components, their broadening
parameters and relative velocity positions were allowed to vary 
in order to
optimize the fit and thus minimize $\chi^2$.
The $\chi^2$ curve as a function of $\Delta v$ in the vicinity of the
global minimum must be smooth and have a parabola-like shape.
The most probable value of $\Delta v$ corresponds to the minimum of
$\chi^2$. In our case it is $-180$ \ms\ for the 16-component model. 
The $1\sigma$ confidence interval
to this value is given
by $\Delta \chi^2 = \chi^2 - \chi^2_{\rm min} = 1$ 
(e.g., Press et al. 1992), 
which is shown by the horizontal dashed line in Fig.~4. 
It is seen from the figure that 
the $1\sigma$ errors are equal to 80 \ms, 65 \ms, and 85 \ms\
for the models in question. 
Adding quadratically the wavelength scale calibration error between
the blue and red arms of
30 \ms\ and using (1), one can easily obtain
\daa = $5.4\pm2.5$ ppm for the main 16-component model.

\begin{figure}[t]
\vspace{0.0cm}
\hspace{-5.0cm}\psfig{figure=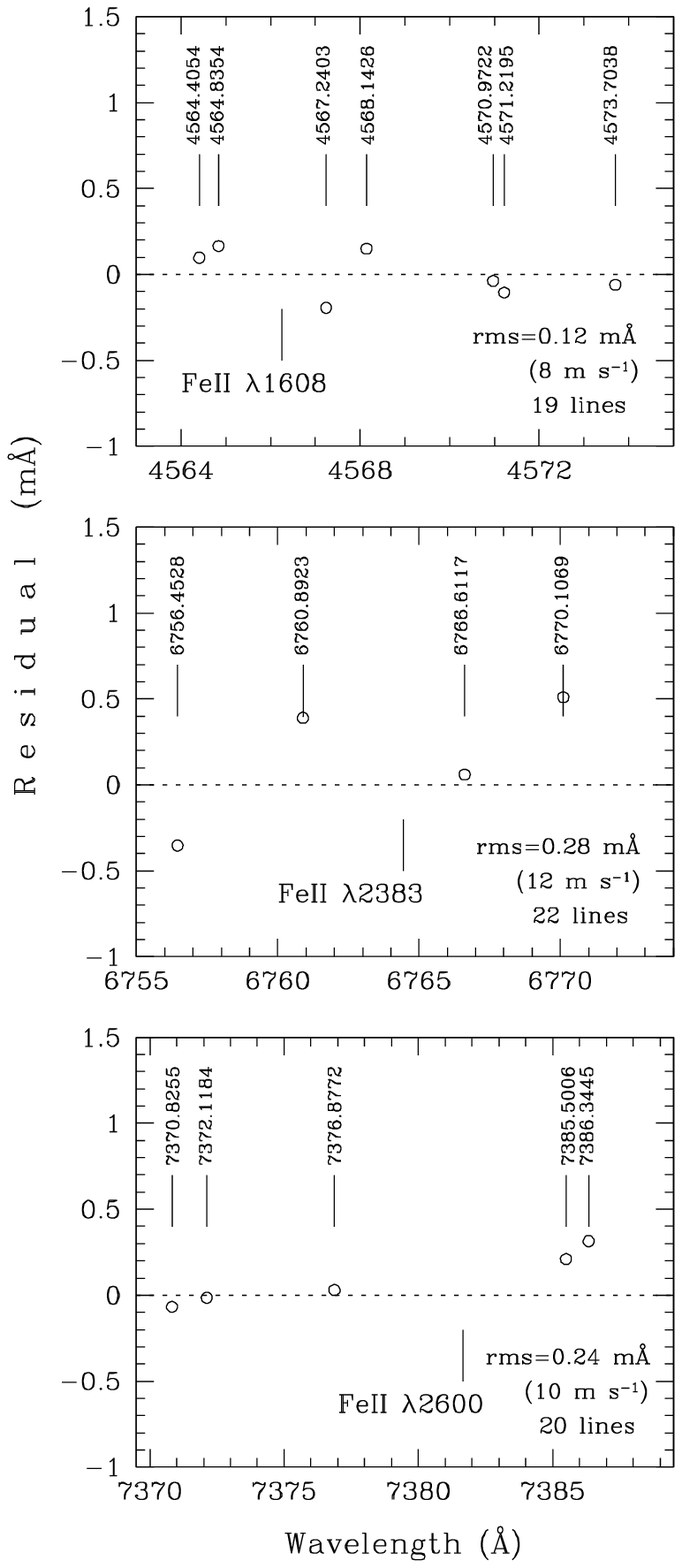,height=19.0cm,width=22cm}
\vspace{-1.5cm}
\caption[]{
An example of the residuals (open circles)
between the laboratory and fitted ThAr wavelengths
for the 4th exposure. The total number of the well-defined
ThAr lines used to calculate the rms error in the echelle order
is indicated on each panel,
whereas only the ThAr lines close to the observed positions (marked by
vertical bars)  of the 
\ion{Fe}{ii} lines of interest are depicted.
}
\label{fig5}
\end{figure}

\begin{table*}[t!]
\centering
\caption{SIDAM analysis: positions, $v$ (in \kms), of the \ion{Fe}{ii}
sub-components,
the line broadening parameters, $b$ (in \kms), and the column densities
$N_{12}$ (in the units of $10^{12}$ \cm) for the 16-, 11-, and
10-component models shown in Figs.~1, 2, and 3
(the corresponding uncertainties of the fitting parameters
are less than 5\%).
The $v$ values are referred to the $\lambda1608$ \AA\ line, whereas
the positions of the  $\lambda2382$ and  $\lambda2600$ \AA\ lines
are shifted at $v' = v + \Delta v$ with the optimal values of
$\Delta v$ given in Fig.~4.
}
\label{tbl-2}
\begin{tabular}{cccccccccc}
\hline
\noalign{\smallskip}
 &
{\footnotesize $v$} & {\footnotesize $b$} & {\footnotesize $N_{12}$}&
{\footnotesize $v$} & {\footnotesize $b$} & {\footnotesize $N_{12}$}&
{\footnotesize $v$} & {\footnotesize $b$} & {\footnotesize $N_{12}$}\\[-2pt]
{\footnotesize No.} &
\multicolumn{3}{c}{\footnotesize 16-component model} &
\multicolumn{3}{c}{\footnotesize 11-component model} &
\multicolumn{3}{c}{\footnotesize 10-component model} \\
\hline
\noalign{\smallskip}
\noalign{\smallskip}
{\scriptsize 1}&
{\scriptsize -74.494}&{\scriptsize 5.36}&{\scriptsize 0.48}\\[-2pt]
{\scriptsize 2}&
{\scriptsize -68.198}&{\scriptsize 1.59}&{\scriptsize 0.11}\\[-2pt]
{\scriptsize 3}&
{\scriptsize -62.067}&{\scriptsize 2.42}&{\scriptsize 0.34}\\[-2pt]
{\scriptsize 4}&
{\scriptsize -39.630}&{\scriptsize 6.39}&{\scriptsize 3.92}&
{\scriptsize -39.643}&{\scriptsize 6.30}&{\scriptsize 3.95}&
{\scriptsize -39.661}&{\scriptsize 6.30}&{\scriptsize 3.91}\\[-2pt]
{\scriptsize 5}&
{\scriptsize -37.029}&{\scriptsize 2.85}&{\scriptsize 2.36}&
{\scriptsize -37.018}&{\scriptsize 2.79}&{\scriptsize 2.30}&
{\scriptsize -37.046}&{\scriptsize 2.82}&{\scriptsize 2.33}\\[-2pt]
{\scriptsize 6}&
{\scriptsize -24.643}&{\scriptsize 7.02}&{\scriptsize 2.90}&
{\scriptsize -24.650}&{\scriptsize 7.14}&{\scriptsize 2.95}&
{\scriptsize -24.497}&{\scriptsize 7.30}&{\scriptsize 3.02}\\[-2pt]
{\scriptsize 7}&
{\scriptsize -17.879}&{\scriptsize 2.05}&{\scriptsize 0.99}&
{\scriptsize -17.837}&{\scriptsize 2.11}&{\scriptsize 1.03}&
{\scriptsize -17.902}&{\scriptsize 1.86}&{\scriptsize 0.90}\\[-2pt]
{\scriptsize 8}&
{\scriptsize -8.875}&{\scriptsize 5.69}&{\scriptsize 2.02}&
{\scriptsize -9.677}&{\scriptsize 4.67}&{\scriptsize 1.58}&
{\scriptsize -8.667}&{\scriptsize 6.13}&{\scriptsize 2.11}\\[-2pt]
{\scriptsize 9}&
{\scriptsize -1.279}&{\scriptsize 3.60}&{\scriptsize 9.56}&
{\scriptsize -1.171}&{\scriptsize 3.88}&{\scriptsize 10.60}&
{\scriptsize -1.072}&{\scriptsize 3.76}&{\scriptsize 10.15}\\[-2pt]
{\scriptsize 10}&
{\scriptsize 4.085}&{\scriptsize 3.31}&{\scriptsize 2.67}&
{\scriptsize 4.607}&{\scriptsize 3.03}&{\scriptsize 2.00}&
{\scriptsize 4.639}&{\scriptsize 2.92}&{\scriptsize 2.01}\\[-2pt]
{\scriptsize 11}&
{\scriptsize 15.493}&{\scriptsize 2.77}&{\scriptsize 0.29}&
{\scriptsize 15.504}&{\scriptsize 3.02}&{\scriptsize 0.31}&
{\scriptsize 16.905}&{\scriptsize 4.47}&{\scriptsize 0.47}\\[-2pt]
{\scriptsize 12}&
{\scriptsize 20.110}&{\scriptsize 1.18}&{\scriptsize 0.28}&
{\scriptsize 20.352}&{\scriptsize 1.50}&{\scriptsize 0.34}\\[-2pt]
{\scriptsize 13}&
{\scriptsize 25.089}&{\scriptsize 3.03}&{\scriptsize 2.31}&
{\scriptsize 25.145}&{\scriptsize 2.89}&{\scriptsize 2.23}&
{\scriptsize 24.873}&{\scriptsize 3.45}&{\scriptsize 2.43}\\[-2pt]
{\scriptsize 14}&
{\scriptsize 35.192}&{\scriptsize 2.95}&{\scriptsize 2.25}&
{\scriptsize 35.232}&{\scriptsize 3.10}&{\scriptsize 2.31}&
{\scriptsize 35.288}&{\scriptsize 2.95}&{\scriptsize 2.27}\\[-2pt]
{\scriptsize 15}&
{\scriptsize 42.746}&{\scriptsize 2.36}&{\scriptsize 0.33}\\[-2pt]
{\scriptsize 16}&
{\scriptsize 47.827}&{\scriptsize 2.31}&{\scriptsize 0.48}\\
\noalign{\smallskip}
\hline
\end{tabular}
\end{table*}

\section{Possible systematics}

We have specifically investigated those effects which might introduce a 
non-zero difference between the blue and the red lines and thus
simulate a \daa\ variation at the ppm level.
In addition to the
systematics discussed in Sect.~2, the following factors should be
taken into account. 

{\it Systematic shift between the blue and red arms.}\
Different velocity offsets may occur in the blue and red frames causing
an artificial Doppler shift between the $\lambda1608$ and $\lambda2382/2600$
lines.  To probe this effect
we used the same procedure as for the QSO exposures and calibrated in
wavelength the spectra of the ThAr arcs. For the blue and red frames
we selected well-exposed ThAr emissions which bracket the positions
of the \ion{Fe}{ii} $\lambda1608$ and $\lambda2382/2600$ lines. 
The calibrated ThAr spectra do not show the rms errors larger than 20 \ms. 
In Fig.~3, an example of the residuals (open circles)
between the laboratory and fitted ThAr wavelengths
for the 4th exposure is shown. The total number of
ThAr lines used to calculate the rms error in the echelle order
is indicated on each panel,
whereas only the ThAr lines close to the observed positions of the 
\ion{Fe}{ii} lines of interest are depicted.
A conservative shift between the blue and red arms of 30 \ms can be obtained
if the rms error of 20 \ms is taken for each arm.

{\it Centering the target onto the slit.}\, 
Radial velocity shifts induced by different changes
of isophote onto the slit are not chromatic, i.e. the shifts are the same in the
blue and red UVES arms. The fact that we have re-centered the object in the
different nights and also before and after passage to the meridian
assure that the effect was different from one exposure to the other,
and, hence, averaged out in the mean spectrum.
Besides, tracking movements act in smoothing the isophote onto the slit.
We note that in our previous results on HE 0515--4414 (Paper~II) we 
also used \ion{Fe}{ii} lines falling in both UVES arms  
without showing evidence of shifts.

{\it Isotopic shifts.}\,
Using the mass shift constants
$k_{\rm MS}$ calculated by Kozlov et al. (2004), we estimated the
uncertainty of \daa~$\simeq 0.77$ ppm caused by unknown isotope abundances
(Paper I) which is equivalent to the uncertainty of about 20 \ms.

{\it Unresolved components.}\,
The normalized residuals,
$({\cal F}^{cal}_i-{\cal F}^{obs}_i)/\sigma_i$,
in Figs.~1, 2, and 3 do not show series which could suggest unresolved
\ion{Fe}{ii} components.  

{\it Possible blends.}\,
The $\lambda1608$ and $\lambda2382$ lines lie in the spectral regions free
from telluric absorptions. No metal absorption lines from other systems
which could blend with these iron lines were found as well.
Weak night sky absorptions affect
the $\lambda2600$ 
profile at $v = -30$ \kms and in the range $v < -50$ \kms which was not
included in the fitting of this line.
Since both the $\lambda2382$ and $\lambda2600$ lines are considered as
having the same radial velocity and the line $\lambda2382$ is clean, this 
contamination of $\lambda2600$ does not influence the final value of \daa.

\section{Conclusions}

We have re-observed the quasar \object{Q 1101--264} with 
spectral resolution $FWHM \sim 3.8$ \kms
and re-analyzed the \ion{Fe}{ii} profiles
associated with the $z = 1.84$ damped Ly$\alpha$ system.
The data represent one of very few spectra of QSOs obtained with
this high resolution and S/N~$\ga 100$.
The newly obtained value of the relative radial velocity shift 
$\Delta v = -180\pm85$ \ms\ between the $\lambda1608$ and $\lambda\lambda2382, 2600$ lines 
represents a factor of 1.5 improvement with respect to
lower resolution archive data ($FWHM \sim 6$ \kms, Paper~I).
Admittedly, there might be further hidden systematic effects which would
challenge the interpretation as due to variation of $\alpha$. If real, it would
correspond to \daa = $5.4\pm2.5$ ppm.  
Thus, the highest spectral resolution achievable on QSO spectra at
8-10~m telescopes is proved to significantly contribute to higher
accuracy in the \daa\ measurement.

We note that averaging the 8 systems in the redshift interval
$1.5 < z < 2.0$ of Chand et al. (2004) one obtains
\daa = $2.4\pm1.2$ ppm. 
On the other hand, Murphy et al. (2003) results
averaged over the same redshift interval (16 systems) give 
\daa = $-5.4\pm4.5$ ppm.
Whereas the Chand et al. and our values are consistent within
the $1\sigma$ confidence interval, 
the value of Murphy et al. is significantly
different.

Up to now, we have obtained high precision measurements of the \daa
with the SIDAM method  
at two redshifts, $z = 1.15$ and $z = 1.84$, only. 
Of course, to probe 
the true redshift evolution of $\alpha$ many more measurements
are needed. It should be understood, however, that such
measurements are extremely hard to perform since they require
spectral data of highest quality
which can be obtained for the brightest QSOs only.
In spite of the fact that many giant telescopes were put into
operation in the last decades, the unique 
simultaneous wavelength coverage
3000 \AA\ $< \lambda <$ 10000 \AA\ of UVES makes it the only
instrument suitable for this work. Thus, the progress in
our understanding the evolution of \daa\ and, hence, of 
dark energy  is expected to be `slow-rolling'
until more powerful Extremely Large Telescopes 
will come into operation.
 
\begin{acknowledgements}
S.A.L. and I.I.A. gratefully acknowledge the hospitality 
of Hamburger Sternwarte.
This research has been supported by
the RFBR grant No.~06-02-16489, 
by the Federal Agency for Science and Innovations
grant NSh~9879.2006.2,
and by the DFG project RE 353/48-1.
P.M. thanks Ville de Paris for an international fellowship.
S.L. was partly supported by the Chilean {\sl Centro de Astrof\'isica}
FONDAP No. 15010003, and by FONDECYT grant N$^{\rm o} 1060823$.

\end{acknowledgements}

\end{document}